\documentstyle[12pt,epsf]{article}

\input epsf

\newcommand{\be}{\begin{eqnarray}}

\newcommand{\ee}{\end{eqnarray}}

\newcommand{\nn}{\nonumber\\ }
\def\labe{\label}

\def\simge{\mathrel{%
   \rlap{\raise 0.511ex \hbox{$>$}}{\lower 0.511ex \hbox{$\sim$}}}}
\def\simle{\mathrel{
   \rlap{\raise 0.511ex \hbox{$<$}}{\lower 0.511ex \hbox{$\sim$}}}}
\def\bigs{\mathrel{
   \rlap{\raise 0.531ex \hbox{$>$}}{\lower 0.531ex \hbox{$<$}}}}

\def\grad{\nabla}                               % gradient
\def\del{\partial}                              % synonym for \partial

\def\frac#1#2{{#1 \over #2}}

\def\half{\ifinner {\scriptstyle {1 \over 2}}
   \else {1 \over 2} \fi}

\def\simge{\mathrel{%
   \rlap{\raise 0.511ex \hbox{$>$}}{\lower 0.511ex \hbox{$\sim$}}}}
\def\simle{\mathrel{
   \rlap{\raise 0.511ex \hbox{$<$}}{\lower 0.511ex \hbox{$\sim$}}}}
\def\bigs{\mathrel{
   \rlap{\raise 0.531ex \hbox{$>$}}{\lower 0.531ex \hbox{$<$}}}}

\def\slashchar#1{\setbox0=\hbox{$#1$}           % set a box for #1 
   \dimen0=\wd0                                 % and get its size
   \setbox1=\hbox{/} \dimen1=\wd1               % get size of /
   \ifdim\dimen0>\dimen1                        % #1 is bigger
      \rlap{\hbox to \dimen0{\hfil/\hfil}}      % so center / in box
      #1                                        % and print #1
   \else                                        % / is bigger
      \rlap{\hbox to \dimen1{\hfil$#1$\hfil}}   % so center #1
      /                                         % and print /
   \fi}                                         %

\def\subrightarrow#1{%                          % #1 under arrow
  \setbox0=\hbox{%                              % set a box
    $\displaystyle\mathop{}%                    % no mathop
    \limits_{#1}$}%                             % just limits
  \dimen0=\wd0%                                 % get width
  \advance \dimen0 by .5em%                     % add a bit
  \mathrel{%                                    % space like =
    \mathop{\hbox to \dimen0{\rightarrowfill}}% % arrow to width
       \limits_{#1}}}                           % text below

\begin{document}
\begin{titlepage}
%\vspace*{1.2cm}

\begin{flushright}
BNL-NT-01/3\\
Saclay-T01/021\\hep-ph/0102009
\end{flushright}
\vspace*{1.2cm}
\begin{center}
{\Large{\bf The Renormalization Group Equation for the Color
 Glass Condensate}}
\vskip0.3cm
 Edmond Iancu,\footnote{CNRS fellow. E-mail: eiancu@cea.fr}\\
{\small\it Service de Physique Th\'eorique\footnote{Laboratoire 
de la Direction des
Sciences de la Mati\`ere du Commissariat \`a l'Energie
Atomique}, CEA Saclay, 91191 Gif-sur-Yvette, France}
        
Andrei Leonidov\footnote{E-mail: leonidov@lpi.ru }\\
         {\small\it P. N. Lebedev Physical Institute,
          Moscow, Russia} 

Larry McLerran\footnote{E-mail: mclerran@quark.phy.bnl.gov }\\
       {\small\it Physics Department, Brookhaven National Laboratory,
                 Upton, NY 11979, USA  }

\end{center}

\date{\today}
\vskip 1.2cm

\parindent=20pt

\begin{abstract}

We present an explicit and simple form of  
the renormalization group equation which governs
the quantum evolution of the effective theory for the 
Color Glass Condensate (CGC). This is a 
functional Fokker-Planck equation for the probability density
of the color field which describes the CGC in the covariant gauge.
It is equivalent to the Euclidean time evolution 
equation for a second quantized current-current
Hamiltonian in two spatial dimensions.
The quantum corrections are included in the leading log approximation,
but the equation is fully non-linear with respect to the
generally strong background field.
In the weak field limit, it reduces to the BFKL equation,
while in the general non-linear case it generates the evolution
equations for Wilson-line operators previously derived by
Balitsky and Kovchegov within perturbative QCD.

\end{abstract}
\end{titlepage}

\setcounter{equation}{0}

The problem of the gross properties of hadron interactions
in the high energy limit has been a central problem of particle and 
nuclear physics for over fifty years.  Remarkable progress in recent 
years has resulted from the observation that the gluon density of
the small $x$ part of the hadron wavefunction, the relevant piece for high 
energies, grows rapidly as $x$ decreases \cite{BFKL,HERA}.
When the gluon density increases,
important non-linear phenomena are expected, which should
eventually lead to {\it saturation} \cite{GLR,MQ,MV94,AM2}, that is, to
a limitation of the maximum gluon phase-space density. At saturation, 
this density is expected to be of order $1/\alpha_s\/$ \cite{GLR,JKMW97,KM98}, 
since interactions among the gluons will cutoff further 
growth once the interaction energy is comparable to the kinetic energy.
For such a large density, perturbation theory breaks down
even if the coupling constant is small, because of strong
non-linear effects.

On dimensional grounds, the square of the
typical momentum scale associated 
with saturation is expected to be of the order of the density 
of gluons per unit area \cite{MV94}.
If this saturation momentum $Q_s$ is large enough ($Q_s \gg
\Lambda_{QCD}$), then the saturation regime is characterized
by weak coupling $\alpha_s(Q_s^2) \ll 1$
and large occupation numbers $\sim 1/\alpha_s$ which,
by the correspondence principle, is a classical regime.
This motivated McLerran and Venugopalan to describe the gluon component
of the hadronic wavefunction at small $x$ as a stochastic
classical Yang-Mills field \cite{MV94}, subsequently
interpreted as a ``Color Glass Condensate'' (CGC) \cite{PI}.
This has the advantage that the non-linear physics at saturation 
can be studied in the simpler setting of a classical
field theory, via a combination of analytic \cite{MV94,KMW,K96,JKMW97,KM98}
and numerical methods \cite{KV00}.

In the McLerran-Venugopalan (MV) model, 
all the dynamical information about the ``fast'' partons 
(i.e., the partons with longitudinal momenta $p^+$ larger then the
scale $\Lambda^+ = xP^+$ of interest\footnote{Throughout, we consider
the hadron in its infinite momentum frame, and use light-cone vector 
notations,  $v^\mu=(v^+,v^-,{\bf v}_\perp)$, with
$v^+\equiv (1/\sqrt 2)(v^0+v^3)$,
$v^-\equiv (1/\sqrt 2)(v^0-v^3)$, and ${\bf v}_\perp
\equiv (v^1,v^2)$. The hadron four-momentum
reads $P^\mu=(P^+,0,{\bf O}_\perp)$, with $P^+\to \infty$,
while soft gluons have $k^+=xP^+$, with $x\ll 1$.}) 
is encoded in a probability distribution 
for the classical color field describing the ``soft''
gluons (i.e., the gluons with momenta $k^+\simle \Lambda^+$).
In principle, this probability distribution can be computed in
perturbation theory, by integrating out the fast quantum modes
in layers of $p^+$, to leading order in $\alpha_s\ln(1/x)$, 
but to all orders in the strong classical background 
field \cite{JKMW97}. This results in a functional renormalization
group equation (RGE) which describes the evolution of the
probability density with the separation scale $\Lambda^+$.
The formal structure of this equation has been first presented
in Refs. \cite{JKLW97}, and further discussed in Refs. \cite{JKW99,PI},
but previous attempts \cite{JKW99} to compute the coefficients 
in this equation have suffered from technical complications
and, moreover, appear to be inconsistent \cite{KMW00}
with the results obtained from perturbative QCD \cite{B,K}.

In a parallel development due to Balitsky \cite{B} and Kovchegov
\cite{K}, the evolution of Wilson line operators has  been
studied within the operator product 
expansion and the dipole model of Mueller \cite{AM3}.  
This has led to a set of coupled
evolution equations for the expectation values of these operators
\cite{B}, which decouple only in the large $N$ limit \cite{K}
with $N$ the number of colors.
(See also Refs. \cite{LT99} for some recent attempts towards solving these
equations.) In a recent paper \cite{W}, Weigert has 
shown that Balitsky's equations can be summarized as a functional 
evolution equation for the generating functional of the Wilson line operators.
As we shall show in this paper, this is precisely the equation
which governs the evolution of the probability distribution for the CGC.

Specifically, our main purpose in this Letter is to present
explicit expressions for the coefficients in the RGE describing
the quantum evolution of the CGC  \cite{JKLW97}.
The details of our calculation, which relies heavily on the formalism
developed in Ref. \cite{PI}, will be presented somewhere else \cite{PII}.

We start by briefly reviewing the MV model
and its quantum evolution (see \cite{PI} for a recent
comprehensive discussion). In the infinite-momentum frame,
and to the order of interest, the fast 
($p^+ >\Lambda^+$) partons can be replaced
by a classical color source, with color charge density $\rho^a(x)$, moving 
along the light-cone ($z=t$ or $x^-=0$) in the positive $x^+$ direction.
Via the Yang-Mills equations:
\be
(D_{\nu} F^{\nu \mu})_a(x)\, =\, \delta^{\mu +} \rho_a(x)\,,
\labe{cleq0}
\ee
 this source determines the classical 
color field $A^\mu_a $ which describes
the soft ($k^+ \simle \Lambda^+$) gluons.
In the saturation regime, the source and the fields
are strong, $A^i\sim \rho \sim 1/g$, and the classical problem
is fully non-linear.

The structure of $\rho^a(x)$ is largerly determined by 
kinematics: Since generated
by shorth-wavelength modes, with momenta $p^+ >\Lambda^+$,
$\rho$  is localized near $x^-=0$,
within a distance $\Delta x^- \simle 1/\Lambda^+$; it is static
(i.e., independent of the
LC time $x^+$) since the soft gluons have a comparatively short
lifetime $\Delta x^+ \sim 1/k^- \propto k^+ \propto x$, during
which the dynamics of the ``fast'' partons can be ignored;
it is a random quantity, since this is the instantaneous color charge 
of the fast partons ``seen'' by the shortlived soft 
gluons at some arbitrary time. The spatial correlators of
the classical source $\rho_a(\vec x)$
(with ${\vec x}\equiv (x^-,{\bf x}_{\perp})$) are inherited
from the (generally time-dependent) quantum correlations of the
fast gluons, via the renormalization group equation to
be presented below. They are encoded into a
weight function for $\rho_a(\vec x)$, denoted as $W_\Lambda[\rho]$, 
which is assumed to be gauge-invariant, and
which depends upon the separation scale $\Lambda^+$ (see below).

Physical quantities like the gluon distribution function
are most directly related to correlations functions of the
transverse fields $A^i$ in the light-cone (LC) gauge $A^+_a=0$.
In the MV model, the correlations functions at the
scale $\Lambda^+$ are obtained as:
\be\labe{clascorr}
\langle A^i_a(x^+,\vec x)A^j_b(x^+,\vec y)
\cdots\rangle_\Lambda\,=\,
\int {\cal D}\rho\,\,W_\Lambda[\rho]\,{\cal A}_a^i({\vec x})
{\cal A}_b^j({\vec y})\cdots\,,\ee
where ${\cal A}_a^i\equiv {\cal A}_a^i[\rho]$ is the 
solution to eq.~(\ref{cleq0}) in the LC gauge, and is 
implicitely given by the following equations
(the other components of the solution vanish: ${\cal A}^+= 
{\cal A}^-=0$):
\be\labe{Atilde}\,
{\cal A}^i\,
(\vec x) &=&{i \over g}\, U(\vec x) \,\partial^i  U^\dagger(\vec x),\nn
U^{\dagger}(x^-,x_{\perp})&=&
 {\rm P} \exp
 \left \{
ig \int_{-\infty}^{x^-} dz^-\,{\alpha}^a(z^-,x_{\perp}) T^a
 \right \},\nn
- \nabla^2_\perp \alpha({\vec x})&=&U^{\dagger}(\vec x)
\, \rho(\vec x) \,U(\vec x) \,\equiv\,{\tilde \rho}(\vec x)\,.
\ee
There are a few important remarks about this classical field:\\
{\it a}) The equations (\ref{cleq0})--(\ref{clascorr})
above are those for a glass (here, a {\it color} glass): 
There is a source, and the source is averaged over. 
This is entirely analogous to what is done for
spin glasses when one averages over background magnetic fields
\cite{PS79}.\\
{\it b}) The solution ${\cal A}^i(x)$ is a two-dimensional pure-gauge
(in the sense that ${\cal F}^{ij}=0$), but not also
a four-dimensional pure-gauge, since the associated
electric field ${\cal F}^{+i}=\partial^+ {\cal A}^i$ is 
non-vanishing. In fact, ${\cal A}^i(x)$ is the
gauge transform with gauge function $U ({\vec x})$
of the solution $\tilde{\cal A}^\mu$
in the covariant gauge $\partial^\mu \tilde A_\mu =0$ 
(COV-gauge), which reads simply
$\tilde{\cal A}^\mu = \delta^{\mu +} \alpha$ with 
$\alpha(\vec x)$ satisfying:
\be\labe{EQTA}
- \nabla^2_\perp \alpha({\vec x})\,=\,{\tilde \rho}(\vec x)\,,
\ee
where ${\tilde \rho}(\vec x)$ is the color source in the COV-gauge.\\
{\it c}) Eq.~(\ref{Atilde}) expresses the LC-gauge solution
${\cal A}^i$ as an {\it explicit} functional of the COV-gauge
source $\tilde \rho\,$:
${\cal A}_a^i\equiv {\cal A}_a^i[\tilde\rho]$.
On the other hand, its dependence upon the LC-gauge source
$\rho$ is not known explicitly. This, 
together with the gauge-invariance of the weight function
$W_\Lambda[\rho]$, suggests that it is more convenient
to perform the functional averaging in eq.~(\ref{clascorr})
by integrating over $\tilde \rho$ (rather than $\rho$).\\
{\it d}) In writing eq.~(\ref{Atilde}), we have chosen
``retarded'' boundary conditions in $x^-$ : 
${\cal A}^i(\tilde x)\to 0$ for $x^- \to -\infty\,$. \\
{\it e)} We shall see later that the source $\tilde \rho$
(and therefore also the ``Coulomb'' field $\alpha$)
has support at $0\simle x^-\simle 1/\Lambda^+$. Thus, when probed by
fields with a poor longitudinal resolution (i.e., which carry
momenta $k^+\ll \Lambda^+$), the configuration
(\ref{Atilde}) appears effectively as a $\theta$-function
in $x^-$ :
\be\labe{APM}
{\cal A}^i(x^-,x_\perp)\,\approx\,\theta(x^-)\,
\frac{i}{g}\,V(\del^i V^\dagger)
\,\equiv\,\theta(x^-){\cal A}^i_\infty(x_\perp),\ee
with ${\cal A}^i_\infty(x_\perp)$ the asymptotic field, and
\be\labe{v}
V^\dagger(x_{\perp})\,\equiv\,{\rm P} \exp
 \left \{
ig \int_{-\infty}^{\infty} dx^-\,\alpha (x^-,x_{\perp})
 \right \},\ee
where the integral over $x^-$ runs effectively up to
$x^-_{max} \simeq 1/\Lambda^+$ (since $\alpha$
vanishes for larger values of $x^-$).
On the same resolution scale, the electric field appears 
as a $\delta$--function:
\be\labe{FDELTA}
{\cal F}^{+i}(\vec x) \,\equiv\,
\partial^+{\cal A}^i\,\approx\,\delta(x^-)\,
{\cal A}^{i}_\infty(x_\perp).\ee

In the original MV model, eqs.~(\ref{cleq0})--(\ref{clascorr}) are simply
postulated, and the weight function $W_\Lambda[\rho]$ is not
computed from first principles.  
Indeed, the physical origin of the large gluon densities,
namely the quantum evolution towards small $x$, is absent from the
model. But in Refs. \cite{JKLW97,JKW99,PI},
one has shown that the MV model is indeed consistent
with the quantum evolution, which entails a 
renormalization group equation (RGE) for the running of $W_\Lambda[\rho]$
with $\Lambda$. This is the same quantum evolution 
which in the weak field regime leads to the BFKL equation.
But for small enough $x$, when one is close to saturation, 
this evolution takes place in the strong background
field ${\cal A}^i$ generated at the previous steps. When ${\cal A}^i\sim 1/g$,
the mean field effects are non-perturbative, and must be included exactly.

To describe the quantum evolution, consider a 
sequence of two classical effective theories (``Theory I''
and ``Theory II'') valid at the scales $\Lambda^+$ and $b\Lambda^+$,
respectively, with $b < 1$ and such as $\alpha_s\ln(1/b) < 1$. 
To the order of interest, the gluon correlations at the lower scale $b\Lambda^+$ 
can be computed in two ways: As classical correlations within Theory II
(which is valid at this scale), or by allowing for quantum fluctuations
in Theory I and integrating out the ``semi-fast'' quantum gluons
with longitudinal momenta in the strip
\be\labe{strip}\,\,
 b\Lambda^+ \,\,<\,\, |p^+|\,\, <\,\,\Lambda^+\,,\ee
to leading order in $\alpha_s\ln(1/b)$, 
but to all orders in the background fields and sources.
(See Refs. \cite{JKLW97,PI} for the quantum generalization of the MV model.)
The difference  $\Delta W
\equiv W_{b\Lambda} - W_\Lambda$, and therefore the evolution
equation for $W_\Lambda$, can be then obtained by
matching these two calculations \cite{PI}. The explicit matching performed
in Ref. \cite{PI} shows that it is indeed
 possible to absorb the effects of the 
semi-fast gluons on the soft correlation functions into a renormalization of
the weight function for the effective theory. This is in agreement with the
original conjecture in  Refs. \cite{JKMW97,JKLW97}, and justifies the use
of the McLerran-Venugopalan model as an effective theory at small $x$.

The RGE is most naturally written in the LC-gauge,
where the quantum effective theory is defined.
It reads\footnote{We use $\tau$ to denote the rapidity
variable, $\tau=\ln(1/x)$,
rather than the more common notation $y$ to avoid confusion
with the space-time coordinate $y$.
Also, we use compact notations, where repeated color indices 
(coordinate variables) are understood to be summed (integrated) over.}
 (with $\tau\equiv\ln(P^+/\Lambda^+)$ and
$d\tau\equiv \ln(1/b)$) \cite{JKLW97,PI} :
\be\labe{RGE}
{\del W_\tau[\rho] \over {\del \tau}}\,=\,\alpha_s
\left\{ {1 \over 2} {\delta^2 \over {\delta
\rho_\tau^a(x_\perp) \delta \rho_\tau^b(y_\perp)}} [W_\tau\chi_{xy}^{ab}] - 
{\delta \over {\delta \rho_\tau^a(x_\perp)}}
[W_\tau\sigma_{x}^a] \right\}\,
\ee
with the coefficients $\sigma_{x}^a\equiv \sigma_a(x_\perp)$ and
$\chi_{xy}^{ab}\equiv\chi_{ab}(x_\perp,y_\perp)$ related to the
1-point and 2-point functions of the color charge $\delta\rho_a(x)$
of the semi-fast gluons via the following relations:
\be\labe{sigperp}
\alpha_s\ln{1\over b}\,\sigma_a ({x}_\perp)&\equiv &
\int dx^- \,\langle\delta \rho_a(x)\rangle\,,\nn
\alpha_s\ln{1\over b}\,\chi_{ab}(x_\perp, y_\perp)&\equiv &
\int dx^- \int dy^-\,
\langle\delta \rho_a(x^+,\vec x)\,
\delta \rho_b(x^+,\vec y)\rangle\,,\ee
where the brackets denote the average over quantum fluctuations
in the background of the tree-level color fields ${\cal A}^i$.

Note that $\sigma$ and $\chi$ are two-dimensional densities
in the transverse plane; the longitudinal structure of the relevant
charge correlators has been integrated over. This is consistent 
with our picture of the quantum evolution where the semi-fast gluons
are unable to discriminate (because of their low $p^+$ momenta)
the longitudinal structure of the source. For this picture
to be validated, however, one still has to check that $\sigma$ and $\chi$
are not sensitive to the internal structure of the source (see below).

There is nevertheless a trace of this longitudinal structure
in eq.~(\ref{RGE}): the functional derivatives there are to be
taken with respect to the color source 
$\rho_\tau^a(x_\perp)\equiv \rho^a(x^- = 1/\Lambda^+,x_\perp)$
in the highest bin of rapidity $(\tau, \tau+d\tau)$,
that is, at $x^-$ in the range $(1/\Lambda^+, 1/b\Lambda^+)$. 
This is so since, as we shall shortly discover,
the quantum corrections have support in that range. 
This longitudinal structure is important to make the
functional derivatives in eq.~(\ref{RGE}) well defined \cite{PI}
(see also eq.~(\ref{DIFFU}) below).

Eq.~(\ref{RGE}) is a functional Fokker-Planck equation 
describing diffusion in the functional space spanned by $\rho$
with ``drift velocity''
$\alpha_s\sigma$ and ``diffusion constant'' $\alpha_s\chi$.
%It can be used to generate ordinary evolution equations
%for $\rho$-dependent observables.
Since $\sigma$ and $\chi$ are generally non-linear functionals of $\rho$
(see below), eq.~(\ref{RGE}) is equivalent with an infinite
set of coupled evolution equations for the correlators of $\rho$.
For these equations to be useful, the functionals $\sigma[\rho]$
and $\chi[\rho]$ have to be known explicitly. To this aim,
it is more convenient to use the COV-gauge source 
$\tilde\rho$ (rather than the LC-gauge source $\rho$)
as the independent variable. Indeed,
$\sigma$ and $\chi$ depend upon the color source
via the classical field ${\cal A}^i$ which is known 
explicitly only in terms of $\tilde\rho$ (cf. eq.~(\ref{Atilde})).

As explained in Ref. \cite{PI}, when going from the LC-gauge to the
COV-gauge, the induced color charge $\sigma$ acquires a new
contribution, in addition to the color rotation with matrix
$U^\dagger$. This is so because the transformation between
the two gauges depends upon the charge content in the problem: this
was $\rho$ at the initial scale $\Lambda^+$, thus giving a gauge
transformation $U^\dagger[\rho]$, but it becomes $\rho+\delta\rho$,
with fluctuating $\delta\rho$, at the new scale $b\Lambda^+$, 
thus inducing a fluctuating component in the corresponding 
gauge function $U^\dagger[\rho+\delta\rho]$.
After averaging out the quantum fluctuations, one is left with a RGE for
$W_\tau[\tilde\rho]$ which is formally similar to eq.~(\ref{RGE}), but
with $\rho \to \tilde\rho$,
$\chi\to \tilde\chi$, and $\sigma\to \tilde\sigma$, where:
\be\label{tildesc}
\tilde \chi_{ab}(x_\perp,y_\perp)&\equiv&
V^{\dagger}_{ac}(x_\perp)\,
\chi_{cd}(x_\perp,y_\perp)\,V_{d b}(y_\perp),\quad\,\,
\tilde\sigma_a (x_\perp)\,\equiv\,
 V^{\dagger}_{ab}(x_\perp)\,\sigma_b(x_\perp)- \delta\sigma_a(x_\perp),
\nn
\delta\sigma_a(x_\perp)&\equiv&
{g\over 2}\,f^{abc}\int d^2y_\perp\,\,
\tilde\chi_{cb}(x_\perp,y_\perp)\,
\langle y_\perp|\,\frac{1}{-\grad^2_\perp}\,|
x_\perp\rangle\,.\ee
The correction $-\delta\sigma$, to be referred to as the
``classical polarization'', is the result of the quantum evolution
of the gauge transformation to the COV-gauge. 

It is our purpose in this Letter to present explicit expressions
for the quantities $\tilde\chi$ and $\tilde\sigma$ as functionals
of $\tilde\rho$, and thus completely specify the RGE for $W_\tau[\tilde\rho]$.
{The Feynman rules for the calculation of these quantities have been
already presented in Refs. \cite{JKLW97,PI}. They involve the propagator 
\be\labe{delAcorr}
iG^{\mu\nu}_{ab}(x,y)[{\cal A}]&\equiv&
\langle {\rm T}\,a_a^\mu(x) a_b^\nu(y)\rangle\ee
of the semi-fast gluons  $a^\mu$
in the presence of the background fields ${\cal A}^i$
and in the LC-gauge $a^+=0$ 
(where the quantum theory is a priori formulated \cite{JKLW97,PI}). 
In Ref. \cite{PI}, this propagator has been constructed
to all orders in the background field, by exploiting the special 
geometry of the latter (see also Refs. \cite{AJMV95,HW98}). 
%More precisely, the propagator is known only away from the
%support of the source (i.e., for $x^- > 1/\Lambda^+$), 
%but this is sufficient for the calculation 
%of $\chi$ and $\sigma$ \cite{PII}. 
One of the subtle points in this construction
%, as discussed in detail in Ref. \cite{PI}, 
has been the choice of a gauge condition (i.e., of an $i\epsilon$ 
prescription for the pole in $1/p^+$ in the gluon propagator)
to fix the residual gauge freedom in the LC-gauge.
This is important since the general results for $\chi$ and $\sigma$
turn out to be dependent upon this 
prescription\footnote{The discrepancy between our results below and those
reported in Refs. \cite{JKW99,KMW00} may be
attributed to using different gauge-fixing prescriptions.}.
}

For consistency with the retarded boundary conditions imposed on
the classical solution (\ref{Atilde}) and with the assumed longitudinal
structure of the original color source (which, we recall, has support
at {\it positive} $x^-$ only, with $0\simle x^-\simle 1/\Lambda^+$),
we shall integrate out the quantum fluctuations by using a
{\it retarded} $i\epsilon$ prescription %for the axial pole %at $p^+=0$
in the LC-gauge propagator (see Ref. \cite{PI} for details).
We have verified that the same results for $\chi$ and $\sigma$
would be eventually obtained also by using an {\it advanced} prescription
\cite{PII}. On the other hand, 
we have not been able to give a sense to calculations with principal value 
or Leibbrandt-Mandelstam prescriptions.

Consider the induced source $\sigma$ first.
Up to some tadpoles (i.e., contributions proportional to $\int d^2p_\perp/
p_\perp^2$) which cancel anyway in the final result, this is obtained from
\be\label{drhoYM}
\langle\delta \rho_a (x)\rangle\,=\,
g f^{abc} \langle(\partial^+ a_{b}^{i}(x) )a_{c}^{i}(x)\rangle
\,=\,
g{\rm Tr}\Bigl(T^a\partial^+_xG^{ii}(x,y)\Bigr)\Big|_{x=y}\,\equiv\,
{\rm Tr}\Bigl(T^a\langle\delta \rho(x)\rangle\Bigl),\ee
where the equal-time limit of the time-ordered propagator
(\ref{delAcorr}) is taken as $y^+=x^++\epsilon$ \cite{PI}.
Eq.~(\ref{drhoYM}) can be evaluated as follows (more details
will be given in Ref. \cite{PII}):
The relevant component of the propagator $G^{ii}(x,y)$
(the only one to give a non-vanishing contribution to
$\langle\delta \rho\rangle$) can be read off eq.~(6.35) of 
Ref. \cite{PI}, which gives (omitting tadpoles once again):
\be\label{rho++}
\langle\delta \rho(x)\rangle\,=\,
-ig\langle x| {\cal D}^i
{1 \over i \partial^+}{\acute G}^{++}  {\cal D}^{\dagger i}|x\rangle,
\ee
where ${\cal D}^i\equiv\partial^i -ig{\cal A}^i_aT^a$ and
${\cal D}^{\dagger j}=\partial^{\dagger j} +ig{\cal A}^j$
with the derivative $\partial^{\dagger j}$ acting on the function on its left.
In eq.~(\ref{rho++}),
$\acute G^{++} (x,y)$ is a component of the background-field
propagator in the {\it temporal} gauge $\acute a^-=0$, 
which enters at intermediate steps in the
construction of the propagator in the LC-gauge \cite{PI}.
It is given by eqs.~(6.22) and (6.13) of Ref. \cite{PI} as:
\be\label{G++c}
{\acute G}^{++}(\vec x,\vec y; p^-)&=&
{2i \over p^-} \int  d^3\vec z\,\,
\partial^i_x G_0(\vec x-\vec z, p^-)\,\delta (z^-)\,
\partial^i_y G_0(\vec z-\vec y, p^-)\nonumber\\
&{}&\,\times\,\left \{
\theta (x^-)\theta (-y^-) V(x_\perp)V^\dagger(z_\perp) -
\theta (-x^-)\theta (y^-) V(z_\perp)V^\dagger (y_\perp)
\right \},
\ee
where  $G_0(p)=1/(2p^+p^--p_\perp^2+i\epsilon)$.
Eq.~(\ref{G++c}) has been written in the $p^-$--representation,
which is convenient given the homogeneity of the problem
in time (recall that the background fields and sources
are independent of $x^+$; cf. eq.~(\ref{Atilde})).
For the same reason, it is preferable to impose the strip
restriction (\ref{strip}) on $p^-$ rather than on $p^+$ \cite{PI} :
\be\labe{strip-}\,
\Lambda^- \,<\, |p^-| \,<\, \Lambda^-/b\,,\ee
where $\Lambda^-\equiv Q_\perp^2/2\Lambda^+$ and $Q_\perp$
is some generic transverse 
momentum\footnote{As explained in \cite{PI}, the two 
restrictions  (\ref{strip}) and (\ref{strip-})
 are equivalent to leading logarithmic accuracy
since the relevant quantum contributions involve nearly on-shell
quanta, with $2p^+p^-\approx p_\perp^2$.}.

The expression in eq.~(\ref{G++c}) has an intuitive 
interpretation \cite{HW98,PI} : The  particle
(a ``semi-fast'' gluon) is moving from $y$ to $x$ in
the presence of the singular electric field (\ref{FDELTA})
located at $z^-=0$. In eq.~(\ref{G++c}), the end points
$y^-$ to $x^-$ lie on opposite sides with respect to the plane
$z^-=0$, which corresponds to trajectories crossing the
potential\footnote{There are, of course, also ``non-crossing''
trajectories where $x^-$ and $y^-$ are of the same sign \cite{PI}.
These have not been included in eq.~(\ref{G++c}) since
they do not contribute to the induced source (\ref{rho++}).}.
Consider $x^->0$ and $y^-<0$ for definitness. On the first
part of the trajectory, from $y^-$ to $z^-=0$, the field ${\cal A}^i$
vanishes (cf. eq.~(\ref{APM}))
and the particle moves freely. At $z^-=0$, the gluon
scatters off the electric field (\ref{FDELTA}) with a scattering
amplitude $V^\dagger(z_\perp)$. Finally, from $z^-=0$ until $x^-$
the color of the gluon is simply rotated, with color matrix $V(x_\perp)$,
by the  ``pure gauge'' background field ${\cal A}^i$.

After inserting eq.~(\ref{G++c}) in eq.~(\ref{rho++}) and
performing simple manipulations, one obtains:
\be\label{drho0}
\langle\delta \rho(x)\rangle&=&g{\cal D}^i_x
\int_{strip} {dp^- \over 2 \pi} {2 \over p^-}\
\int {dp^+ \over 2 \pi} {dk^+ \over 2 \pi}\,
{e^{-i(p^+-k^+)x^-}  \over p^+ + i \epsilon} 
\int d \Gamma_{\perp}
(p_{\perp} \cdot k_{\perp}) G_0(p) G_0(k)
\nn &{}&\qquad
\left \{
\theta(p^-) V(x_\perp)V^\dagger(z_\perp) 
-\theta(-p^-) V(z_\perp)V^\dagger (y_\perp)
\right \}
{\cal D}^{\dagger i}_y {\Big |}_{y=x}\,,
\ee
where it is understood that $k^-=p^-$ (with $p^-$ restricted to
the strip (\ref{strip-})), and we have used the following
shorthand notation for the transverse phase-space integrals:
\be\label{dGamma}
\int d \Gamma_{\perp}\,\equiv\, \int d^2z_{\perp} \int {d^2p_\perp \over (2 \pi)^2}
\int {d^2k_\perp \over (2 \pi)^2}\,\,
{\rm e}^{ip_{\perp}\cdot(x_{\perp}-z_{\perp})}\,
{\rm e}^{ik_{\perp}\cdot(z_{\perp}-y_{\perp})}.\ee
Note the $+ i \epsilon$ prescription for the ``axial'' pole
at $p^+=0$ : this is the retarded gauge-fixing
prescription alluded to before. By using this prescription, together
with the Feynman prescriptions in the propagators $G_0(p)$ and $ G_0(k)$,
the integrals over $p^+$ and $k^+$ can be easily performed via contour
technics. This generates $\theta$-functions of $\pm x^-$ and $\pm p^-$
which, together with the $\theta$-functions already present in
eq.~(\ref{drho0}), have the effect to single out contributions proportional
to $\theta(x^-)\theta(-p^-)$.

The final result of this straightforward calculation reads:
\be\label{sig1x}
\langle\delta \rho_a (x)\rangle\,=\,g\theta(x^-)\,\frac{e^{-ib\Lambda^+ x^-}-
e^{-i\Lambda^+ x^-}}{\pi x^-} \,{\cal D}^i_x
\int d \Gamma_{\perp}\,{p_{\perp} \cdot k_{\perp} \over p_{\perp}^2 k_{\perp}^2}
\,k^i\,{\rm Tr}\Bigl(T^a V_z V^\dagger_y\Bigr)\Big |_{y_{\perp}=x_{\perp}},\ee
where the $x^-$--dependent ``form factor'' has been generated by
integrating over $p^-$ within the strip (\ref{strip-})
(with $ k_{\perp}^2/2\Lambda^-\approx \Lambda^+$ to the accuracy
of interest) :
\be\label{LOGX1}
\int_{strip} {dp^- \over 2 \pi}\,
{\theta(-p^-)\over 2(p^-)^2}\,e^{\,i {k_{\perp}^2 \over 2p^-}x^-}\,=\,
\frac{1}{k_{\perp}^2}\,\frac{e^{-ib\Lambda^+ x^-}-
e^{-i\Lambda^+ x^-}}{2\pi i x^-}\,.\ee
Remarkably, eq.~(\ref{sig1x}) shows that $\langle\delta \rho\rangle$
has support only at positive, and relatively large, $x^-\,$, with
(typically) $1/\Lambda^+ \simle x^- \simle 1/b\Lambda^+\/$.
That is, the induced
source is located on top on the tree-level source $\rho$
(with support at $x^-\simle 1/\Lambda^+$). By 
induction, we conclude that the classical source has support only
at $x^-\ge 0\,$. When the quantum fluctuations are integrated out in layers
of $p^+$, the source receives new contributions in layers of $x^-$, with
essentially no overlap between successive layers.
This specific longitudinal picture is intimately related to our use
of the retarded prescription in the LC-gauge propagator.
With an advanced prescription,
one would rather obtain a source with support at negative $x^-$. 
Still, the {\it integrated} (or two-dimensional)
charge density $\sigma(x_\perp)$ comes up the same with both prescriptions
\cite{PI,PII}, and reads (cf. eq.~(\ref{sigperp}))
\be\label{sigx}
\alpha_s\ln{1\over b}\,\sigma^a(x_\perp) \,=\,{g\over  \pi}\,\ln{1\over b}\,
{\cal D}^i_x 
\int d\Gamma_{\perp}\,{p_{\perp} \cdot k_{\perp} \over p_{\perp}^2 k_{\perp}^2}
\,k^i\,{\rm Tr}\Bigl(T^a V_z V^\dagger_y\Bigr)\Big |_{y_{\perp}=x_{\perp}},\ee
where the covariant derivative involves the asymptotic
field: ${\cal D}^i_x=\partial^i_x -i{\cal A}^i_\infty(x_\perp)$.
Note that the logarithmic enhancement has been generated only after
integration over $x^-$.
For what follows, it is useful to have the expression of 
$\sigma$ rotated to the covariant gauge:
\be\label{tsigx}
gV^{\dagger}_{ab}(x_\perp)\,
\sigma^b(x_\perp) \,=\,{4}\,\partial^i_x
\int d\Gamma_{\perp}\,{p_{\perp} \cdot k_{\perp} \over p_{\perp}^2 k_{\perp}^2}
\,k^i\,{\rm Tr}\Bigl(T^a V^\dagger_x V_z\Bigr).\ee
In the weak field limit $g\alpha \ll 1$,
the r.h.s. of this equation scales like $g\rho$, as it can be easily
verified by expanding the Wilson lines; thus $\sigma = {\cal O}(\rho)$.
In fact, in this limit, $\sigma = {\cal K}_{virt}\rho$, with ${\cal K}_{virt}$
the virtual BFKL kernel \cite{BFKL}. In the saturation regime, 
$g\alpha \sim 1$,
and all the non-linear effects displayed by the above equation are equally
important.

Consider $\chi$ now. A typical contribution to the charge-charge
correlator reads \cite{JKLW97,PI}
\be\labe{chi2}
\langle\delta \rho_a(x)\,
\delta \rho_b(y)\rangle\,=\,4ig^2\,
{\cal F}^{+i}_{ac}(\vec x)\, G^{ij}_{cd}(x,y)\,{\cal F}^{+j}_{db}(\vec y),\ee
where it is understood that $y^+=x^+ +\epsilon$.
The calculation of this and the remaining contributions
is similar to that of $\sigma$ outlined above,
and eventually gives the following result
(in matrix notations) \cite{PII}:
\be\labe{chi1}
\langle\delta \rho_x\,
\delta \rho_y\rangle&=&{g^2\over  \pi}\,\ln{1\over b}\,\Biggl\{{\cal F}^{+i}_x
%\delta^{ij}_\perp(x_{\perp}-y_{\perp})
\langle x_\perp|\delta^{ij}-\frac{\partial^i\partial^j}{\grad^2_\perp}\,
|y_\perp\rangle{\cal F}^{+j}_y\\
&{}&\qquad +\,
\biggl(\rho+2{\cal F}^{+i}\biggl({\cal D}^i - {\partial^i\over 2}\biggr)\biggr)_x
\langle x_\perp|\,\frac{1}{-\grad^2_\perp}\,|y_\perp\rangle\,
\biggl(\rho+2\biggl({\cal D}^{\dagger j} 
- {\partial^{\dagger j}\over 2}\biggr){\cal F}^{+j}\biggr)_y
\Biggr\}.\nonumber\ee
In this equation, both $\rho$ and ${\cal F}^{+i}$ are 
localized at $x^- \simle 1/\Lambda^+$. Thus,
unlike the induced charge (\ref{sig1x}), the charge-charge correlator
appears to be localized near the LC (in both $x^-$ and $y^-$), 
and thus sensitive to the internal structure 
of the source. If this was also true for $\chi$, it would spoil the separation
of scales assumed by the effective theory, and thus the validity of the latter.
Fortunately, however, this is not the case: Like the electric field 
${\cal F}^{+i}=\partial^+{\cal A}^i$, all the other functions 
carrying the $x^-$ and $y^-$ dependences in eq.~(\ref{chi1})
are {\it total derivatives} with respect to $x^-$ or $y^-$. Thus, $\chi$ 
--- which is obtained from  eq.~(\ref{chi1}) by integrating
over $x^-$ and $y^-$, cf. eq.~(\ref{sigperp}) ---
is sensitive only to the asymptotic fields.

Specifically, by using the equations of motion (\ref{cleq0}),
the following identity can be verified (for an arbitrary function
$\Phi(x_\perp)$) :
\be
(\rho+2{\cal F}^{+i}{\cal D}^i)_x\Phi(x_\perp)\,=\,i\partial^+{\cal D}^2_x
\Phi(x_\perp),\ee
which allows us to cast $\chi$ in the following form:
\be\labe{chi}
\chi_{ab}(x_\perp,y_\perp)\,=\,4\,\biggl\{{\cal A}^i_x
\delta^{ij}_\perp(x_{\perp}-y_{\perp}){\cal A}^j_y\,+\,
\Bigl({\cal D}^i{\cal A}^i\Bigr)_x
\langle x_\perp|\,\frac{1}{-\grad^2_\perp}\,|y_\perp\rangle\Bigl({\cal A}^j
{\cal D}^{\dagger j}\Bigr)_y\biggr\}_{ab},\ee
where all the fields are asymptotic fields, e.g., %, cf. eq.~(\ref{APM}); e.g.,
${\cal A}^i_x\equiv {\cal A}^i_\infty(x_\perp)$ and 
${\cal D}^i=\partial^i -ig{\cal A}^i_\infty$. For weak fields,
${\cal A}^i\approx -\partial^i\alpha$, and $\chi\approx \rho\,{\cal K}_{real}\,
\rho\,$, with ${\cal K}_{real}$ the real BFKL kernel \cite{BFKL}.
After rotation to the COV-gauge (cf. eq.~(\ref{tildesc})), eq.~(\ref{chi}) yields:
\be\label{tchi}
\tilde \chi_{ab}(x_\perp,y_\perp)\,=\,\frac{4}{g^2}
\,\biggl\{\partial^iV^{\dagger}_x\,
\delta^{ij}_\perp(x_{\perp}-y_{\perp}) \partial^jV_y\,+\,
\partial^i_x\biggl(\Bigl(\partial^iV^{\dagger}\Bigr)_x
\langle x_\perp|\,\frac{1}{-\grad^2_\perp}\,|y_\perp\rangle
\Bigl(\partial^j V\Bigr)_y\biggr)\partial^{\dagger j}_y\biggr\}_{ab},\,\,\ee
where the derivatives not included in the brackets act on all the
functions on their right (or left).

Once $\tilde \chi$ is known, we are in a position to compute also the classical
polarization $-\delta\sigma$ in eq.~(\ref{tildesc}). One thus obtains
(up to a tadpole which would anyway cancel against the tadpole
ignored when writing eq.~(\ref{tsigx})) :
\be\label{classpol}
-g\delta\sigma_a(x_\perp)
\,=\,{-2i}\,\partial^i_x
\int d\Gamma_{\perp}\,{p_{\perp} \cdot k_{\perp} \over p_{\perp}^2 k_{\perp}^2}
\,\partial^i_x\,{\rm Tr}\Bigl(T^a V^\dagger_x V_z\Bigr).\ee
%\equiv V^{\dagger}_x \chi_{xy} V_y 
Unlike $\sigma$ in eq.~(\ref{tsigx}), $\delta\sigma$ has no contribution linear
in $\rho$, so it does not contribute at the BFKL level. This is as it should,
since the BFKL equation is already obtained by combining the weak field
limits of the previous results for $\sigma$ and $\chi$ \cite{JKLW97}.
On the other hand, $\delta\sigma$ is as important as $\sigma$ in the
non-linear regime where $g\alpha \sim 1$. By combining eqs.~(\ref{tsigx})
and (\ref{classpol}), we finally arrive at the following expression for
$\tilde\sigma$, which is the quantity which enters
the RGE for $W_\tau[\tilde\rho]$ (cf. eq.~(\ref{tildesc})) :
\be\label{tsigma}
g\tilde\sigma_a (x_\perp)\,=\,-\grad_\perp^2\left\{2i\,
\int d\Gamma_{\perp}\,{p_{\perp} \cdot k_{\perp} \over p_{\perp}^2 k_{\perp}^2}
\,{\rm Tr}\Bigl(T^a V^\dagger_x V_z\Bigr)\right\}
\,\equiv\,-g\,\grad_\perp^2\nu_a (x_\perp).\ee
% \langle\delta\alpha_a (x_\perp)\rangle .\ee
Up to a normalization factor, the quantity $\nu_a (x_\perp)$ introduced above
is the induced color field in the COV-gauge, that is, the modification
in the original field $\alpha$ induced by the quantum corrections.
It can be verified that $\nu$, and therefore $\tilde\sigma$, are
free of tadpoles:
the tadpoles hidden in $\sigma$, eq.~(\ref{tsigx}), and $-\delta\sigma$,
eq.~(\ref{classpol}), have cancelled in their sum.

At this stage, the RGE for $W_\tau[\tilde\rho]$ is fully specified.
It has the formal structure shown in eq.~(\ref{RGE}), but
with $\rho \to \tilde\rho$,
$\chi\to \tilde\chi$, and $\sigma\to \tilde\sigma$, and 
$\tilde\chi$ and $\tilde\sigma$ given by eqs.~(\ref{tchi}) and (\ref{tsigma}). 
But the peculiar structure of the ``virtual'' correction
in eq.~(\ref{tsigma}) suggests that the RGE may acquire an even
simpler form by using the COV-gauge field $\alpha$ rather than the
COV-gauge source $\tilde\rho$ as the functional variable of $W_\tau$.
By using (cf. eq.~(\ref{EQTA}))
\be
{\delta \over {\delta \tilde\rho^a_\tau(x_\perp)}}\,=\,\int d^2z_\perp\,
\langle x_\perp|\,\frac{1}{-\grad^2_\perp}\,|z_\perp\rangle\,
{\delta \over {\delta \alpha^a_\tau(z_\perp)}}\,,\ee
where $\alpha^a_\tau(x_\perp)\equiv
\alpha^a(x^- = 1/\Lambda^+,x_\perp)$ is the color field 
in the highest bin of rapidity (where the quantum corrections
are located), one can transform the RGE for $W_\tau[\tilde\rho]$ 
into the following RGE for $W_\tau[\alpha]$ :
\be\labe{RGEA}
{\del W_\tau[\alpha] \over {\del \tau}}\,=\,\alpha_s
\left\{ {1 \over 2} {\delta^2 \over {\delta
\alpha_\tau^a(x_\perp) \delta \alpha_\tau^b(y_\perp)}} 
[W_\tau\eta_{xy}^{ab}] - 
{\delta \over {\delta \alpha_\tau^a(x_\perp)}}
[W_\tau\nu_{x}^a] \right\}\,,
\ee
where 
\be\label{nu}
g\nu^a (x_\perp)&=&
2i\int {d^2z_\perp\over (2\pi)^2}\,\frac{1}{(x_\perp-z_\perp)^2}
\,{\rm Tr}\Bigl(T^a V^\dagger_x V_z\Bigr),
\ee
and 
\be\label{eta}
g^2\eta^{ab}(x_\perp,y_\perp)&\equiv&g^2\int d^2z_\perp \int d^2u_\perp\,
\langle x_\perp|\,\frac{1}{-\grad^2_\perp}\,|z_\perp\rangle\,
\tilde\chi^{ab}(z_\perp,u_\perp)\,
\langle u_\perp|\,\frac{1}{-\grad^2_\perp}\,|y_\perp\rangle\,\nn
&=&4\int {d^2z_\perp\over (2\pi)^2}\,
\frac{(x^i-z^i)(y^i-z^i)}{(x_\perp-z_\perp)^2(y_\perp-z_\perp)^2 }
\Bigl\{1+ V^\dagger_x V_y-V^\dagger_x V_z - V^\dagger_z V_y\Bigr\}^{ab}.\ee
The kernels in the equations above have been written in coordinate space
by using
\be
 \int {d^2p_\perp \over (2 \pi)^2}
\int {d^2k_\perp \over (2 \pi)^2}\,\,
{p_{\perp} \cdot k_{\perp} \over p_{\perp}^2 k_{\perp}^2}\,\,
{\rm e}^{ip_{\perp}\cdot(x_{\perp}-z_{\perp})}\,
{\rm e}^{ik_{\perp}\cdot(z_{\perp}-y_{\perp})}\,=\,{1\over (2\pi)^2}\,
\frac{(x^i-z^i)(y^i-z^i)}{(x_\perp-z_\perp)^2(y_\perp-z_\perp)^2 }\,
.\ee

The functional derivatives with respect to $\alpha_\tau$ in eq.~(\ref{RGEA})
are understood
as variations in the color field at $x^-\sim 1/\Lambda^+$, which is
the end point of the Wilson lines $V^\dagger$ and $V$ (cf. the remark
after eq.~(\ref{v})). Thus, the derivatives of the latter read simply:
\be\label{DIFFU}
{\delta V^\dagger (x_{\perp})\over \delta \alpha^a_\tau(z_\perp)}\,=\,
igT^aV^\dagger (x_{\perp})\,\delta^{(2)}(x_{\perp}-z_\perp),\quad
{\delta V(y_{\perp})\over \delta \alpha^a_\tau(z_\perp)}\,=\,
-igV (y_{\perp})T^a\delta^{(2)}(y_{\perp}-z_\perp).\ee

Eqs.~(\ref{RGEA})--(\ref{eta}) represent our main result in this Letter.
They govern the flow with $\tau=\ln(1/x)$ of the probability density
$W_\tau[\alpha]$ for the stochastic color field $\alpha_\tau(\vec x)$
which describes the CGC in the COV-gauge. Since the r.h.s. of eq.~(\ref{RGEA})
is a total derivative with respect to $\alpha$, 
this flow automatically preserves the
correct normalization of the weight function: 
\be\label{norm}
\int {\cal D}\alpha\, \,W_\tau[\alpha]\,=\,1.\ee
%at any $\tau$. 

The functional RGE equation
(\ref{RGEA}) can be used to derive evolution equations for all the
observables which can be related to $\alpha$
(like the soft gluon correlation functions (\ref{clascorr})). 
If $O(\alpha)$ is any such an operator,
then its average over $\alpha$, defined as in eq.~(\ref{clascorr})
but with $\rho$ replaced by $\alpha$, obeys the following equation
\be\labe{evolO}
{\del \over {\del \tau}}\langle O(\alpha)\rangle_\tau\,=\,\alpha_s\,
\left\langle {1 \over 2}\,\eta_{xy}^{ab}\, {\delta^2  O\over {\delta
\alpha_\tau^a(x_\perp) \delta \alpha_\tau^b(y_\perp)}}\, + \,\nu_{x}^a\,
{\delta O\over {\delta \alpha_\tau^a(x_\perp)}}\,\right\rangle_\tau,
\ee
which is obtained by multiplying eq.~(\ref{RGEA}) with
$O(\alpha)$, integrating over $\alpha$, and performing
some integrations by parts in the r.h.s.

In particular, by chosing $O(\alpha)={\rm tr}(V^\dagger_x V_y)$
with the Wilson lines $V^\dagger_x$ and $V_y$ in the 
{\it fundamental} representation, and by using eqs.~(\ref{nu})--(\ref{DIFFU})
and performing the color algebra,
one eventually obtains the following evolution equation
(with $N$ the number of colors)
\be\labe{evolV}
{\del \over {\del \tau}}\langle {\rm tr}(V^\dagger_x V_y)
\rangle_\tau={\alpha_s\over 2 \pi^2}\int d^2z
\frac{(x_\perp-y_\perp)^2}{(x_\perp-z_\perp)^2(y_\perp-z_\perp)^2 }
\left\langle {\rm tr}(V^\dagger_x V_z)\,{\rm tr}(V^\dagger_z V_y)
- N {\rm tr}(V^\dagger_x V_y)\right\rangle_\tau,\,\,\ee
which coincides with the corresponding equation obtained by Balitsky \cite{B}
within perturbative QCD. In the large $N$ limit, the 4-point function
in the r.h.s. factorizes:
\be
\left\langle {\rm tr}(V^\dagger_x V_z)\,{\rm tr}(V^\dagger_z V_y)\right\rangle_\tau
\longrightarrow 
\left\langle {\rm tr}(V^\dagger_x V_z)\right\rangle_\tau\,
\left\langle{\rm tr}(V^\dagger_z V_y)\right\rangle_\tau\quad({\rm for}\,\,N\to\infty),
\ee 
and eq.~(\ref{evolV}) reduces a closed equation for the quantity
${\cal N}(x_\perp,y_\perp)\equiv 
\langle{\rm tr}(1-V^\dagger_x V_y)\rangle_\tau$ (which
represents the forward scattering amplitude of a color dipole 
off the hadron): 
\be\labe{evolN}
{\del \over {\del \tau}} {\cal N}_{xy}
={\alpha_s\over 2 \pi^2}\int d^2z
\frac{(x_\perp-y_\perp)^2}{(x_\perp-z_\perp)^2(y_\perp-z_\perp)^2 } 
\left\{ N\Bigl({\cal N}_{xz} + {\cal N}_{zy} - {\cal N}_{xy}\Bigr)
- {\cal N}_{xz}{\cal N}_{zy}\right\}.\ee 
This coincides with the evolution equation obtained by Kovchegov \cite{K}
within the framework of Mueller's dipole model \cite{AM3}.

We now return to more formal properties of the RGE (\ref{RGEA}), and notice
the following remarkable relation between the coefficients in this equation:
\be
{1 \over 2} \int d^2y \,{\delta\eta^{ab}(x_\perp,y_\perp) \over  
\delta \alpha^b_\tau(y_\perp)}
\,=\,\nu^a(x_\perp)\,,\ee
which is most probably a consequence of gauge symmetry
(as it relies 1-point and 2-point functions). By using this relation,
the RGE can be brought into a Hamiltonian form:
\be\labe{RGEH}
{\del W_\tau[\alpha] \over {\del \tau}}\,=\,{\alpha_s \over 2}\int d^2x\int d^2y\,
{\delta \over {\delta
\alpha_\tau^a(x)} }\left(\eta_{xy}^{ab}\,
{\delta W_\tau\over {\delta \alpha_\tau^b(y)}}
\right)\,\equiv\,-\,H W_\tau, 
\ee
with the following, manifestly positive definite, Hamiltonian
\be\labe{H}
H&=&\int {d^2 z_\perp\over 2\pi }\,J^i_a(z_\perp)\,J^i_a(z_\perp),\nn
J^i_a(z_\perp)&\equiv& \int {d^2 x_\perp\over 2\pi }\,
\frac{z^i-x^i}{(z_\perp-x_\perp)^2}\,(1 - V^\dagger_zV_x)_{ab}\,
{i \delta \over {\delta
\alpha_\tau^b(x_\perp)} }.\ee
(Notice that the ``current'' operator $J^i_a(z_\perp)$ 
is indeed Hermitian, since the Wilson lines in the
adjoint representation are real color matrices: $V^\dagger_{ab} = V_{ba}$.)
This same Hamiltonian structure has been recently identified by
Weigert \cite{W} in relation with a generating functional for
Balitsky's evolution equations. Since our effective theory generates
Balitsky's equations indeed, it should not be a surprise that we come 
across the same Hamiltonian.

Eq.~(\ref{RGEH}) has the structure of a diffusion equation. 
Since the Hamiltonian is a purely kinetic term, this
equation describes a Brownian motion in the functional space of $\alpha$.
As $\tau$ increases, we expect the solution $W_\tau[\alpha]$ to this equation 
to spread over the whole available phase space, and thus to
go asymptotically to zero at any given ``point'' $\alpha^a(x_\perp)$
(because of the normalization condition (\ref{norm})). 
As a simple analogy, consider the equation describing
the one-dimensional diffusion:
\be
{\del W_\tau(x) \over {\del \tau}}\,=\,{1 \over 2}\,\eta\,
{\del^2 W_\tau(x) \over {\del x^2}}\,.\ee
The solution to this equation, 
\be W_\tau(x)\,=\,{1\over \sqrt{2\pi\eta \tau}}\,
{\rm e}^{-\frac{x^2}{2\eta\tau}},\ee
is normalized to unity at any $\tau$
($\int dx W_\tau(x) =1$), and goes smoothly to zero at any $x$ when
$\tau\to \infty$, as it spreads over the whole $x$ axis.

In Ref. \cite{W}, Weigert has conjectured that a uniform (i.e,
$\alpha$-independent) probability distribution $W_0$ should be a fixed
point of the evolution as $\tau\to \infty$. His arguments are
as follows: first, any constant $W_0$ is an eigenfunction of
$H$ with zero eigenvalue ($HW_0=0$); second, since $H$ is positive definite,
all its other eigenvalues must be positive.
From this, he concludes that the constant eigenfunction (the
``fundamental state'' of $H$) will be singled
out by the evolution for sufficiently large $\tau$.
However, we disagree with this interpretation. As mentioned above, a uniform 
probability density is necessarily trivial, $W_0=0$,  to cope with
the normalization condition. Besides, for the ``fundamental state'' of $H$ 
to be singled out by the evolution, there must be a gap
between this state and the first excited state. But such a gap
cannot be expected for a Hamiltonian which is just a kinetic term, 
like eq.~(\ref{H}). 
To distinguish between these various asymptotic scenarios and, even
more importantly, to find the behaviour of $W_\tau[\alpha]$ at large,
but finite $\tau$, further studies of eq.~(\ref{RGEH}) are necessary.

To conclude, we have established the equivalence between the
non-linear evolution equations derived within perturbative QCD and
within the effective theory for the Color Glass Condensate.
We have thus resolved a discrepancy between these two approaches
previously reported in Refs. \cite{JKW99,KMW00}. There are several
noticeable differences between our calculations and those in Refs. 
\cite{JKW99,KMW00} (chiefly among them, the different gauge-fixing 
prescriptions),  which may explain the different results, 
but a deep understanding of these differences requires more study. 
We have found that
the evolution of the gluon correlation functions in the non-linear
regime at small $x$ is governed by a Hamiltonian describing
current-current interactions in two spatial dimensions. 
Given the relative simplicity of this Hamiltonian,
we believe that analytic solutions to this equation (at least
in particular limits) should be within our reach.

\newpage %\bigskip
{\large{\bf Acknowledgements}}

The authors gratefully acknowledge conversations with 
Ian Balitsky, Jean-Paul Blaizot, Elena Ferreiro,
Jamal Jalilian-Marian, Al Mueller, 
Eugene Levin, Yuri Kovchegov, Alex Kovner, Raju Venugopalan, 
and Heribert Weigert, with special mentions to the latter (H.W.)
for having explained to us some subtle points of his recent work.

This manuscript has been authorized under Contract No. DE-AC02-98H10886 
with the U. S. Department of Energy.

\end{document}